\documentclass[a4paper, 10pt, conference]{IEEEtran}      

\newtheorem{theorem}{\textbf{Theorem}}
\newtheorem{corollary}{\textbf{Corollary}}
\newtheorem{definition}{\textbf{Definition}}
\newtheorem{lemma}{\textbf{Lemma}}
\newcommand{\Rmnum}[1]{\expandafter\@slowromancap\romannumeral #1@}
\makeatother         
\usepackage{amsmath}
\usepackage{amssymb}
\usepackage{graphicx}
\usepackage{epstopdf}
\usepackage{amscd}
\usepackage{bm}
\usepackage{subfigure}
\usepackage{cite}
\usepackage{algorithm}
\usepackage{algorithmic}
\def\R{{\mathbb{R}}}
\def\x{{\mathbf{x}}}
\def\L{{\mathcal{L}}}
\def\X{{\mathbf{X}}}
\def\G{{\mathcal{G}}}
\def\E{{\mathcal{E}}}
\def\V{{\mathcal{V}}}
\def\U{{\mathbf{U}}}

\def\S{{\mathcal{S}}}

\def\T{{\text{T}}}
\def\f{{\text{f}}}
\def\rank{{\operatorname{rank}}}

\graphicspath{{figure/}}

\title{
On Critical Sampling of Time-Vertex Graph Signals
}

\author{\IEEEauthorblockN{Junhao Yu, Xuan Xie, Hui Feng, Bo Hu}\IEEEauthorblockA{\textit{Research Center of Smart Networks and Systems, School of Information Science and Technology, Fudan University} \\Shanghai 200433, China \\\{jhyu17, xxie15, hfeng, bohu\}@fudan.edu.cn}}

\begin{document}

\maketitle
\thispagestyle{empty}
\pagestyle{empty}

\begin{abstract}
Joint time-vertex graph signals are pervasive in real-world. This paper focuses on the fundamental problem of sampling and reconstruction of joint time-vertex graph signals. We prove the existence and the necessary condition of a critical sampling set using minimum number of samples in time and graph domain respectively. The theory proposed in this paper suggests to assign heterogeneous sampling pattern for each node in a network under the constraint of minimum resources. An efficient algorithm is also provided to construct a critical sampling set. 
\end{abstract}

\section{INTRODUCTION}

Sampling theory of graph signal aims to recover the whole signal by using part of the observation of the original signal, which can save the cost to infer in a large graph. Various methods have been developed to reconstruct the original signal from noise-free samples\cite{chen2015discrete,marques2015sampling}, or noisy observations\cite{xie2017design,anis2016efficient,tsitsvero2016signals,chamon2017greedy,8683739,sakiyama2019eigendecomposition}, based on bandlimitedness or smoothness prior in graph spectral domain.


Most related works focused on the static graph signal. But many real-world signals are time-varying, like the temperatures collected by a sensor network, which means the signal on each vertex is of a higher dimensional form like a vector or tensor. In such cases, the joint time-vertex graph signal is a candidate model to describe and process such kind of signals whose frequency spectrum can be obtained by so-called \textit{Joint Time-Vertex Fourier Transform} (JFT)\cite{grassi2018time}. R. Varma \textit{et al.} define the smooth signal on joint time-vertex model and propose a recovery strategy\cite{varma2019smooth}. Besides, Wei \textit{et al.} propose a sampling scheme for continuous time-varying graph signals\cite{wei2019optimal}. Ji \textit{et al.} extend the time domain to Hilbert space and introduce a generalized graph signal processing framework\cite{8646656}.

In this paper, we investigate the fundamental sampling theory, i.e, the conditions for critical sampling, for joint time-vertex graph signals in noise-free scene. Some prior works have touched this problem. From the view of product graphs,  {Ortiz-Jim{\'e}nez} \textit{et al.} extend the bandlimited signal to the simultaneously bandlimited (SBL) signal and propose a sampling scheme in two domains separately \cite{ortiz2018sampling}. The generalized graph signal processing theory\cite{8646656} discusses some properties of sampling. 
However, they don't propose the scheme of critical sampling with minimum samples, which we will show later in section \ref{sec:sampling}.


In this paper, we reveal the connection between general bandlimited signal (GBL) and simultaneously bandlimited signal (SBL) on the time-vertex graph by introducing the projection bandwidth. Then, we give the necessary conditions for critical sampling on GBL signal in two domains. Finally, we propose an algorithm to find a critical sampling set, which is proved to exist.

\section{MODEL}

\subsection{Graph Signal and Sampling Theory}

Consider an undirected graph $\mathcal{G}=(\mathcal{V},\mathcal{E},\mathbf{W})$ with the set of vertex $\mathcal{V}$, edge $\mathcal{E}$ and weighted adjacency matrix $\mathbf{W}$. A graph signal is $\mathbf{x}=[x_1,x_2,\dots,x_N]$ in which the element $x_i$ represents the signal value at the $i$-th vertex in $\mathcal{V}$.

The graph Laplacian is $\L=\mathbf{D}-\mathbf{W}$, where the degree matrix $\mathbf{D}=\text{diag}(\mathbf{1W})$. Because $\L$ is symmetric, it has the spectral decomposition
\begin{equation}
\L=\mathbf{U}\Lambda \mathbf{U}^H,
\end{equation}
where the eigenvectors $\{\mathbf{u}_i\}_{i=1}^N$ of $\L$ form the columns of $\mathbf{U}$, and $\Lambda$ is a diagonal matrix of eigenvalues $\{\lambda_i \}_{i=1}^N$  according to $\{\mathbf{u}_i\}$. The eigenvalues can be regarded as frequencies and eigenvectors can be regarded as Fourier-like basis for graph signals\cite{ortega2018graph}. The Graph Fourier Transform (GFT) can be represented by $\mathbf{x}_\text{f}= \mathbf{U}^H \x$ and Inverse Graph Fourier Transform (IGFT) can be represented by $\x=\U\x_\f$. In this sense, a graph signal $\x$ is so-called bandlimited signal when $\x_\f$ has $K<N$ non-zero coefficients, which has the low-dimensional representation as 
\begin{equation}
\label{equ:low_dimension}
\x=\tilde{\U}\tilde{\x}_\f,
\end{equation}
where $\tilde{\x}_\f$ consists of non-zero spectral components in $\x_\f$, and $\tilde{\U}$ is constructed by extracting the columns of $\U$ corresponding to the indices of the non-zero elements of $\x_\f$\cite{ortega2018graph, chamon2017greedy}. 

Define the sampled graph signal $\x_\S=[x_{s_1},\dots,x_{s_M}]$, such that $\x_\S=\Psi\x$, where $\S=\{s_1,\dots, s_M\}$ is the index set of sampled vertices, and the sampling matrix $\Psi\in\{0,1\}^{M\times N}$ is defined as
\begin{equation}
[\Psi]_{i,j}=\begin{cases}
1,& j=s_i;\\
0,& \text{otherwise}.
\end{cases}
\end{equation}
The interpolation matrix $\Phi$ is the operator of recovering $\x_\S$ to $\x'=\Phi\x_\S \in \R^N$. The following Theorem \ref{thm:chen}  gives the condition of perfect reconstructing $\x$ from $\x_\S$\cite{chen2015discrete}.

\begin{theorem}
\label{thm:chen}
Define $\tilde{\U}_M=\Psi \tilde{\U}$, for all bandlimited graph signal $\x$ with bandwidth $K$. If $\Psi$ satisfies $\rank(\tilde{\U}_M)=K$, perfect recovery $\x=\Phi \Psi \x$ can be achieved by choosing $\Phi=\tilde{\U}(\tilde{\U}_M^\T\tilde{\U}_M)^{-1}\tilde{\U}_M^\T$.
\end{theorem}

Obviously, the rank condition of Theorem \ref{thm:chen} is necessary for perfect reconstruction as the following corollary. 

\begin{corollary}
\label{cor:thm1}
If there exists a linear interpolation operator to recovering $\x$ from $\x_\S$, there must be $\rank(\tilde{\U}_M)=K$, i.e. we need at least $K$ samples. 
\end{corollary}

We call a sampling matrix $\Psi$ a \textit{qualified sampling matrix} when it satisfies  $\rank(\tilde{\U}_M)=K$. And we call the sampling set $\S$ corresponding to a qualified sampling matrix a \textit{qualified sampling set}.

\subsection{Joint Time-vertex graph signal and Joint Time-vertex Fourier Transform}
Now we consider an undirected graph $\mathcal{G}_G=(\mathcal{V}_G,\mathcal{E}_G,\mathbf{W}_G)$, and each vertex relates to a time sequence of length $T$, which can be represented by a cycle graph $\G_T=(\V_T,\E_T,\mathbf{W}_T)$. A joint time-vertex graph, denoted by $\G_J$, is constructed by Cartesian product of $\G_T$ and $\G_G$ as shown in Fig. \ref{fig:example} \cite{grassi2018time},

\begin{equation}
\label{equ:product graph}
\G_J = \G_T \times \G_G=(\V_T\times \V_G,\E_J).
\end{equation}

\begin{figure}[t]
\centering
\includegraphics[scale=0.4]{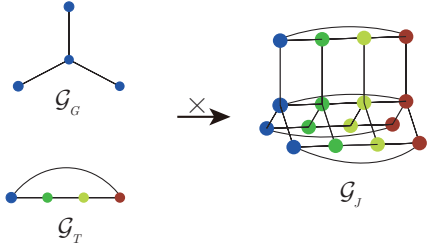}
\caption{A joint time-vertex graph}
\label{fig:product_graph}
\end{figure}

Denoting the graph signal at instant $t$ by $\x_t\in \R^N$, the total graph signal is represented as the matrix $\X=[\x_1, \x_2, \dots, \x_T]\in \R^{N\times T}$ with the corresponding vectorized form $\x=\operatorname{vec}(X)\in \R^{NT}$. 

The Laplacian matrix of $\G_J$, denoted by $\L_J$, is the Cartesian product of the Laplacian of $\G_T$ and $\G_G$,
\begin{align*}
\L_J &= \L_T\times \L_G=(\L_T\otimes I_G)+(I_T\otimes \L_G)\\&= \U_J \Lambda_J \U_J^H 
= (\U_T\otimes \U_G)(\Lambda_T\times\Lambda_G )(\U_T \otimes \U_G)^H,
\end{align*}
where $\otimes$ denotes the Kronecker product and $I_T,I_G$ are the identify matrices which have the same size as $\L_T,\L_G$ \cite{grassi2018time}.

JFT has been introduced by appling Fourier transform of $\G_T$ in time domain and Fourier transform to $\G_G$ in vertex domain \cite{grassi2018time}
\begin{equation}
\X_\f=\text{JFT}\{\X\}=\U_G^H \X \U_T.
\end{equation}
Expressed in vector form, the transform becomes 
\begin{equation}
\x_\f=\text{JFT}\{\x\}=\U_J^H \x.
\end{equation}

\section{Sampling on Joint Time-vertex Graphs}
\label{sec:sampling}
Because the joint time-vertex graph consists of two domains, there are different meanings when we talk about a bandlimited signal.

\begin{definition}\textit{(GBL)}
A joint time-vertex graph signal $\x$ is a GBL signal when $\x_\f$ has $K<(NT)$ none-zero elements, where $K$ is the general bandwidth.
\end{definition}
\begin{definition}\textit{(Projection bandwidth)}
\label{def:projection bandwidth}
For a GBL signal $\X$, when $\X_\f$ has $K_G$ non-zero columns, we define the projection bandwidth on $\G_G$ as $K_G$. And when $\X_\f$ has $K_T$  non-zero rows, we define the projection bandwidth on $G_T$ as $K_T$.
\end{definition}

The projection bandwidth builds the connection of $\G_J$ with $\G_G$ and $\G_T$ respectively. When a GBL signal $\X$ has projection bandwidth $K_G$ on $\G_G$ and $K_T$ on $\G_K$, each column of $\X$ is a bandlimited signal on $\G_G$ with bandwidth $K_G$, and each row of $\X$ is a bandlimited signal on $\G_T$ with bandwidth $K_T$.

\begin{definition}\textit{(SBL)\cite{ortiz2018sampling}}
\label{def:simultaneously}
We call a GBL signal $\X$ an SBL signal, if its projection bandwidth $K_G<N$ and $K_T<T$.
\end{definition}

Obviously, the relationship between projection bandwidth and general bandwidth is
\begin{equation}
\label{equ:relationship}
\text{max}(K_T,K_G)\le K\le K_TK_G.
\end{equation}
So if a signal $\X$ is SBL, it must be GBL. But a GBL signal may not be SBL.
For example, when the spectral coefficient $\X_\f$ is a diagonal matrix with all non-zero diagonal entries, the signal $\X$ is a GBL signal, but it is not an SBL signal.


An SBL signal $\X$ admits a low-dimensional representation as
\begin{equation}
\label{equ:low_dimensional2}
\x=(\tilde{\U}_T \otimes \tilde{\U}_G)\tilde{\x}_\f \Leftrightarrow \X=\tilde{\U}_G\tilde{\X}_\f\tilde{\U}_T^H,
\end{equation} 
where $\tilde{\X_\f}$ and $\tilde{\x}_\f$ are the non-zero spectral components in $\X_\f$ and $\x_\f$. And $\tilde{\U}_T$ and $\tilde{\U}_G$ are obtained by removing the columns of $\U_T$ and $\U_G$ corresponding to the indices of the rows and columns of $\X_\f$ that are all zero.  

Based on Theorem \ref{thm:chen} and $\rank(\tilde{\U}_T \otimes \tilde{\U}_G)=\rank(\tilde{\U}_T)\rank(\tilde{\U}_G)$, a separately sampling scheme of SBL signals is proposed in \cite{ortiz2018sampling}. Let $S_T\subseteq \V_T$ and $\S_G\subseteq \V_G$ be two subset of vertices from $\G_T$ and $\G_G$. There must be a qualified sampling set with $|\S_T|\ge K_T$ and $|\S_G|\ge K_G$ so that we can recover  $\x$ from $\x_\S$, which can be expressed as
\begin{equation}
\label{equ:seperatly}
\X_{\S_G\times\S_T}=\Psi_G \X \Psi_T^\T=\Psi_G \tilde{\U}_G\tilde{\X}_\f\tilde{\U}_H^\T\Psi_T^\T,
\end{equation}
where $\Psi_T$ and $\Psi_G$ are sampling matrices of sampling sets $\S_T$ and $\S_G$.  The vectorized form of $\X_{\S_G\times\S_T}$ can be expressed as  
\begin{equation}
\label{equ:seperatly2}
\x_{\S_T\S_G}=\left[\Psi_T \tilde{\U}_T \otimes \Psi_G \tilde{\U}_G \right]^H \tilde{\x}_\f.
\end{equation}

In the separate sampling scheme \cite{ortiz2018sampling}, the actual sampling set of $\G_J$ can be denoted by $\S=\S_T\times \S_G$ so that the number of samples is $|\S|=|\S_T||\S_G|$.

But the separate sampling scheme may not give a qualified sampling set with minimum vertices, since it sampled at least $K_TK_G$ vertices\cite{ortiz2018sampling}. Applying Theorem \ref{thm:chen} to $\G_J$, for all GBL signals with general bandwidth $K$ and projection bandwidth $K_T$ and $K_G$, there will always exist a qualified sampling set of $\G_J$, denoted by $\S$, satisfying $|\S|= K$. If we hope to squeeze the sample size from $K_TK_G$ to $K$, we need to analyze this question from the view of the joint time-vertex rather than considering it separately.

Before presenting our main theorem, we first define the projection set on graphs. As the vertex set of $\G_J$ in Eq. (\ref{equ:product graph}) is $\V_T\times \V_G$, these vertices can be represented as a two-tuple form like $(1,1), (1,2), \dots, (T,N)$. 
\begin{definition}\textit{(Projection set of sampling set on two graphs)}
Given a sampling set $\S\subset \V_T \times \V_G$, we define the projection set on $\V_T$ and $\V_G$ as $S_T$ and $\S_G$, respectively, where $|\S_T|$ means how many time-slots we need to sample at least on one node, and $|\S_G|$ means how many vertices of $\G_G$ we need to sample during all the time. 
\end{definition}

For example, $\S=\{(1,2),(2,2),(3,4)\}$, then $\S_T=(1,2,3)$ and $\S_G=(2,4)$. The projection sets on two graphs would reveal additional bounds of a qualified sampling set of $\G_J$. Besides the rank condition from Corollary \ref{cor:thm1}, we are interested in whether there are any additional conditions of qualified sampling set. Before proposing the theorem, we prove a lemma first.

\begin{lemma}
\label{lem:sample}
For a bandlimted signal $\x$ with bandwidth $K$ on a graph $\G=(\V,\E)$, there are two sampling sets of the signal $\x$, denoted as $S_1$ and $S_2$. When $S_1 \subseteq S_2$, if $S_2$ is not a qualified sampling set, $S_1$ is not a qualified sampling set either.
\end{lemma}

\begin{IEEEproof}
Denote the sampling matrix of $\S_1$ and $\S_2$ by $\Psi_1$ and $\Psi_2$. When $\S_1 \subseteq \S_2$, $\rank(\Psi_1\tilde{\U})\le \rank(\Psi_2\tilde{\U})$. Because $\S_2$ is not a qualified sampling set, from Corollary \ref{cor:thm1}, we can conclude $\rank(\Psi_2\tilde{\U})<K$, $\rank(\Psi_1\tilde{\U})<K$. So $\S_1$ is not a qualified sampling set.
\end{IEEEproof}

\begin{theorem}
\label{thm:sample}
For any GBL signal $\x$ on $\G_J$ with general bandwidth $K$ and projection bandwidth $K_T$ and $K_G$, if $\S$ is a qualified sampling set of $\G_J$, i.e. its corresponding sampling matrix $\Psi$ satisfies $\rank(\Psi \tilde{\U}_J)=K$, there must be:

\begin{enumerate}
\item $|\S|\ge K$
\item $|\S_G|\ge K_G$
\item $|\S_T|\ge K_T$
\end{enumerate}
\end{theorem}

\begin{IEEEproof}
\label{pro:proof_th2}
$|\S|\ge K$ is obvious by applying Corollary \ref{cor:thm1} to $\G_J$. We prove clause 2) by contradiction. And clause 3) can be proved in the same way. 

 Assume there is a sampling set $\S$ whose projection sampling set on $\G_G$ satisfies $|\S_G|<K_G$. We construct another sampling set $\S'=\V_T\times \S_G$. The sampled signal on $\S'$ is denoted by $\X_{\S'}$. Now recovering the original signal $\X$ from $\X_{\S'}$ is equivalent to recovering each column of $\X$ from the corresponding column of $\X_{\S'}$. It means that a sampling set with $|\S_G|$ vertices is a qualified sampling set of a bandlimited signal with bandwidth $K_G$, which is not possible according to Corollary \ref{cor:thm1}. So $\S'$ is not a qualified sampling set. Since $\S\subset \S'$, from Lemma \ref{lem:sample}, $\S$ is not a qualified sampling set either. So there must be $|\S_G|\ge K_G$.
\end{IEEEproof}

\begin{definition} \textit{(Critical sampling set)}
A qualified sampling set $\S$ is a \textit{critical sampling set} on $\G_J$, when it satisfies $|\S|=K$, $|\S_T|=K_T$ and $|\S_G|=K_G$ at the same time.
\end{definition}
The corresponding sampling matrix $\Psi$ of a critical sampling set is called \textit{critical sampling matrix}. A critical sampling leads to the minimum cost in many scenes. For example, a critical sampling set of a sensor network signal means we can use as less as possible sensors, time-slots and data to recover the whole signal. 

Regarding the existence of critical sampling set and critical sampling matrix, we have the following corollary.

\begin{corollary}
\label{cor:exists}
For any GBL signals, there always exists a critical sampling matrix and its corresponding sampling set. 
\end{corollary}

\begin{IEEEproof}
Consider a GBL signal $\X$ with vectorized form $\x$, with general bandwidth $K$ and the projection bandwidth $K_T$ and $K_G$. According to Eq. (\ref{equ:low_dimension}) and (\ref{equ:low_dimensional2}), we have $\tilde{\U}_J$, $\tilde{\U}_T$ and $\tilde{\U}_G$. Define $\tilde{\U}_J'=\tilde{\U}_T\otimes \tilde{\U}_G$ and we have $\rank(\tilde{\U}_J')=\rank(\tilde{\U}_T)\rank(\tilde{\U}_G)=K_TK_G$. Using separately sampling scheme by Eq. (\ref{equ:seperatly2}), we get a qualified sampling matrix $\Psi_T$ on $\G_T$ and a qualified sampling matrix $\Psi_G$ on $\G_G$. Then we define $\Psi' = \Psi_T\otimes \Psi_G$. The corresponding sampling set $\S'$ of $\Psi'$ has projection sampling set $\S'_T$ and $\S'_G$ satisfying $|\S'_T|=K_T$ and $|\S'_G|=K_G$. Obviously $\rank(\Psi'\tilde{\U}'_J)=K_TK_G$. 

If $K=K_TK_G$, $\Psi'$ is a critical sampling matrix of $\X$. If $K<K_TK_G$, the column set of $\tilde{\U}_J$ is a subset of the column set of $\tilde{\U}'_J$. So the column set of $\Psi'\tilde{\U}_J$ is a subset of $\Psi'\tilde{\U}'_J$. Now $\Psi'\tilde{\U}_J \in \R^{K_TK_G \times K}$ and $\rank(\Psi'\tilde{\U}_J)=K$.  There always exists a sampling matrix $\Psi_c \in \{0,1\}^{K\times K_TK_G}$ such that $\rank(\Psi_c\Psi'\tilde{\U}_J)=K$. Let $\Psi=\Psi_c\Psi'$. Since $\S'$ satisfies $|\S'_T|=K_T$ and $|\S'_G|=K_G$, the corresponding sampling set $\S$ of $\Psi$ satisfies $|\S_{T}|\le K_T$ and $|\S_{G}|\le K_G$. Because $\rank(\Psi\tilde{\U}_J)=K$, $\Psi$ is a qualified sampling matrix, such that $|\S_{T}|\ge K_T$, $|\S_{G}|\ge K_G$, by Theorem \ref{thm:sample}. Now we get  $|\S_{T}|= K_T$, $|\S_G|= K_G$. As $\Psi\in \{0,1\}^{K\times NT}$, $|\S|=K$. So $\Psi$ is the critical sampling matrix. 
\end{IEEEproof}

According to our proof of Corollary \ref{cor:exists}, we propose an efficient algorithm (Algorithm \ref{alg:find}) to find a critical sampling set. Provided a corresponding sampling matrix $\Psi$ from $\S$, we can get the original signal $\x$ by interpolation matrix $\Phi=\tilde{\U}_J(\Psi \tilde{\U}_J)^{-1}$.

Algorithm \ref{alg:find}\footnote{An example code is showed on https://github.com/ParaNoth/Example-code-of-On-Critical-Sampling-of-Time-Vertex-Graph-Signals} provides a feasible way to find the critical sampling set and reduces the time complexity compared with the algorithm proposed in \cite{chen2015discrete}. For example, we can use Gaussian Elimination to find the index set of maximal whose time complexity is $O(N^3)$ when the matrix has $N$ rows. So the time complexity of the algorithm based \cite{chen2015discrete} is $O((NT)^3)$ because $\tilde{\U}_J$ has $NT$ rows, while the time complexity of our algorithm is $O(N^3)+O(T^3)$(step 1 in Algorithm \ref{alg:find}) and $O((K_TK_G)^3)$(step 3 in Algorithm \ref{alg:find}). 

\begin{algorithm}[htbp] 
\caption{Finding a critical sampling set}
\label{alg:find}
\begin{algorithmic}[1]
\REQUIRE $\tilde{\U}_T$, $\tilde{\U}_G$, $\tilde{\U}_J$
\ENSURE $\S$
\STATE Find $\S_T$, $\S_G$, the index set of maximal linearly independent rows of $\tilde{\U}_T$, $\tilde{\U}_G$, respectively.
\STATE Choose the rows of $\tilde{\U}_J$ based on $\S'=\S_T\times \S_G$, and then get $\Psi'\tilde{\U}_{J}$.
\STATE Get $\S$ from maximal linearly independent rows of $\Psi'\tilde{\U}_{J}$.

\end{algorithmic}
\end{algorithm}

\section{Example}
\label{sec:example}
In this section, we show an example of joint time-vertex graph as Fig. \ref{fig:product_graph} to explain our idea.  The Laplacian matrices of two undirected graphs $\G_T,\G_G$ are
$$
\L_T=\left[\begin{matrix}
2&-1&0&-1\\
-1&2&-1&0\\
0&-1&2&-1\\
-1&0&-1&2\\
\end{matrix} \right], \L_G=\left[\begin{matrix}
1&-1&0&0\\
-1&3&-1&-1\\
0&-1&1&0\\
0&-1&0&1\\
\end{matrix} \right].
$$
The GBL graph signal $\X$ on graph $\G_J$ with $K=3$, $K_T=2$, $K_G=2$ is as
$$
\X=\left[\begin{matrix}
0.2985&-0.3533&-0.2985&0.3533\\
0&0&0&0\\
-0.1492&0.5432&0.1492&-0.5432\\
-0.1492&-0.1898&0.1492&0.1898\\
\end{matrix} \right], 
$$
whose corresponding frequency coefficient is
$$
\X_\f=\left[\begin{matrix}
0&0&0&0\\
0&0.733&0&0\\
0&0.612&0.517&0\\
0&0&0&0\\
\end{matrix} \right].
$$
So $\tilde{\U}_T$ and $\tilde{\U}_G$ are
\begin{equation*}
\tilde{\U}_T=\left[\begin{matrix}
0&0.7071\\
-0.7071&0\\
0&-0.7071\\
0.7071&0\\
\end{matrix} \right], \tilde{\U}_G=\left[\begin{matrix}
0&0.8165\\
0&0\\
-0.7071&-0.4082\\
0.7071&-0.4082\\
\end{matrix} \right].
\end{equation*}

\subsection{Finding a critical sampling set}

We use Algorithm \ref{alg:find} to find a critical sampling set for $\X$. From $\tilde{\U}_T$ and $\tilde{\U}_G$, we get $\S_T=\{1,2\}$ and $\S_G=\{1,3\}$ (step 1 in Algorithm \ref{alg:find}), so $\S'=\{(1,1),(1,3),(2,1),(2,3)\}$ as step 2 in Algorithm \ref{alg:find}. We have
\begin{equation}
\Psi'\tilde{\U}_J=\left[
\begin{matrix}
0&0&0.5774\\
0&0&-0.2887\\
0&-0.5774&0\\
0.5&0.2887&0\\
\end{matrix}
 \right].
\end{equation}

\begin{figure}[htbp]
\centering    
\includegraphics[scale=0.4]{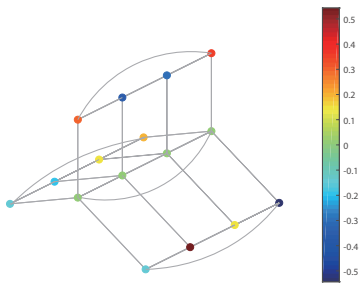}   
\caption{Original signal in section \ref{sec:example}}
\label{fig:example} 
\end{figure}

By Gaussian elimination, we can get $\S=\{(1,1), (2,1), (2,3)\}$ (step 3 in Algorithm \ref{alg:find}). The original signal is shown in Fig. \ref{fig:example} and the critical sampling set is shown in Fig. \ref{fig:qualified}(a). Now $\S$ satisfies $\S=3$, $\S_T=2$ and $\S_G=2$, so it is the critical sampling set. Compared with separately sampling scheme, we sampled $3$ vertices which less than $K_TK_G=4$. 

\subsection{Substitution between time and vertex}
In many scenes, the sampling cost in time and vertices are different, so there might be a trade-off between time and vertices. For example, in a sensor network, sensors with low-speed ADC are cheap, while sensors with high-speed ADC may be much more expensive. Is it possible to use more sensors in exchange of lower sampling frequency? If so, is there any limit of mutual substitution between sampling in time and vertices? Theorem \ref{thm:sample} actually answers the questions and gives the bound of the substitution, which means we can substitute between time and vertices within certain limits. 

\begin{figure}[htbp]
\centering    

\subfigure[]
{
	\begin{minipage}{3cm}
	\centering          
	\includegraphics[scale=0.4]{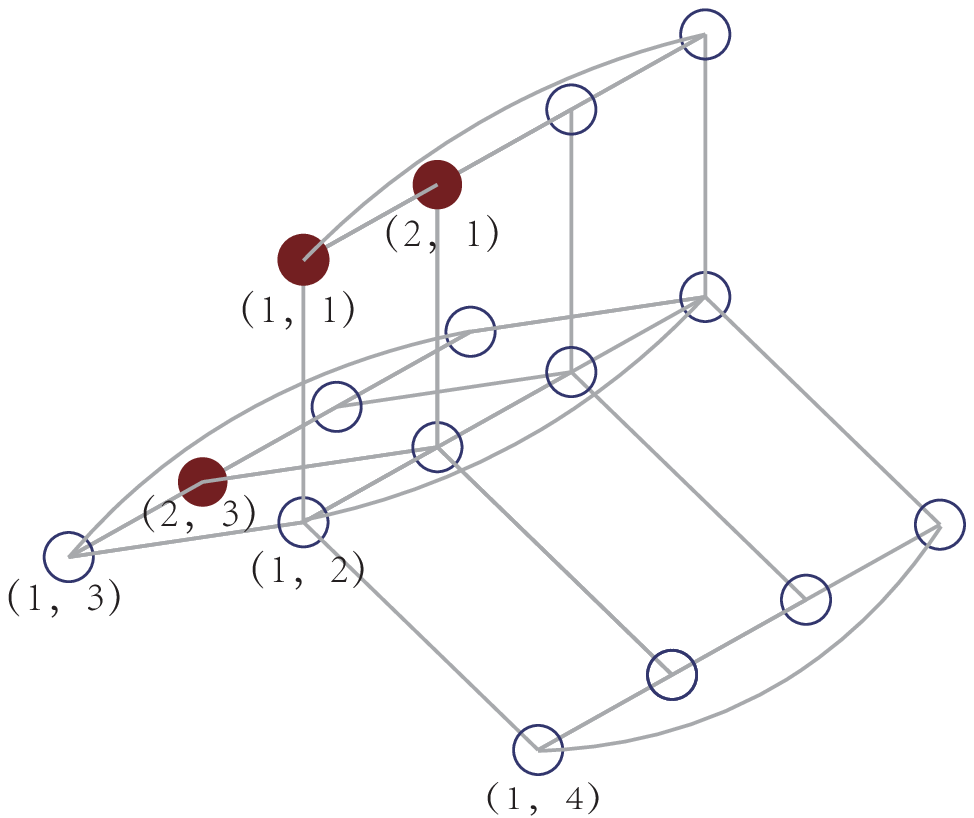}   
	\end{minipage}
}
\hfill
\subfigure[]
{
	\begin{minipage}{4cm}
	\centering      
	\includegraphics[scale=0.4]{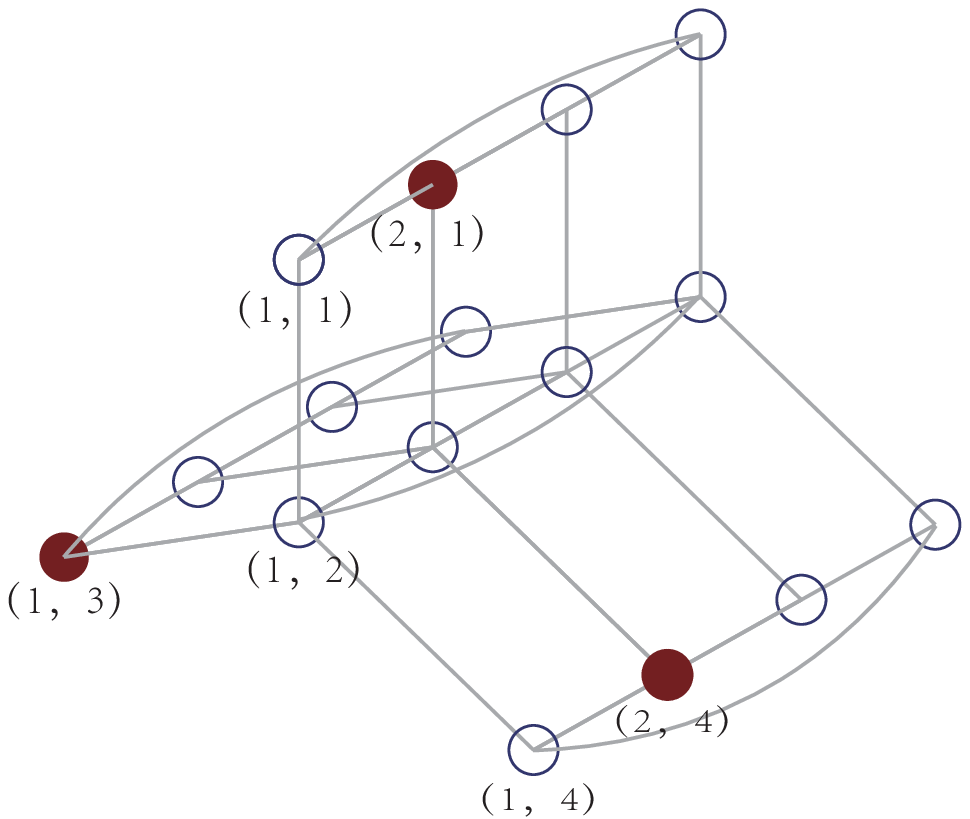}   
	\end{minipage}
}

\caption{Two qualified sets, (a) is the critical sampling set and (b) has lower sampling frequency}
\label{fig:qualified} 
\end{figure}

For example, there are two qualified sampling sets, shown in Fig. \ref{fig:qualified}, but only Fig. \ref{fig:qualified}(a) is a critical sampling set. Compared to Fig.  \ref{fig:qualified}(a), Fig. \ref{fig:qualified}(b) opens the sensor 4 in order to reduce the sampling frequency of sensor 1. Conversely, Fig. \ref{fig:qualified}(a)  increases the sampling frequency of sensor 2 so that we can close the sensor 4. But we can not close any more sensors. Otherwise, we cannot recover the original signal. 

Fig. \ref{fig:qualified}(a) also reveals that when a signal is a GBL signal, there might be a qualified set with different sampling frequency on every node. This property is important for sampling design in sensor networks, social networks, etc.

\section{CONCLUSION}
We have shown that we should sample in joint time-vertex domain rather than sampling in two domain separately, if we want to get a more efficient sampling. The main result of this paper can be extended to all product graph signals. In future works, we plan to investigate the continuous as time-varying graph signals.

\section*{ACKNOWLEDGMENT}
This work was supported by the National Key Research and Development Program of China (No. 213), the Shanghai Municipal Natural Science Foundation (No. 19ZR1404700), and the NSF of China (No. 61501124).

\bibliographystyle{IEEEbib}
\bibliography{ref}

\end{document}